\documentclass[preprint2]{emulateapj}
\usepackage{graphicx}
\usepackage{subfigure}

\begin{document}

\title{PROBING TOPOLOGICAL RELATIONS BETWEEN HIGH-DENSITY AND LOW-DENSITY REGIONS OF 2MASS WITH HEXAGON CELLS}

\author{Yongfeng Wu$^{1*}$ and Weike Xiao$^2$}
\affil{$^1$American Physical Society, San Diego, CA, USA; yongfeng.wu@maine.edu \\
$^2$Department of Astronautics Engineering, Harbin Institute of Technology, P.O. Box 345, Heilongjiang Province 150001, China\\
Received 2013 January 27; accepted 2013 November 16; published 2014 January 16
}


\begin{abstract}
We introduced a new two-dimensional (2D) hexagon technique for probing the topological structure of the universe
in which we mapped regions of the sky with high and low galaxy densities onto a 2D lattice of hexagonal unit cells.
We defined filled cells as corresponding to high-density regions and empty cells as corresponding to low-density
regions. The numbers of filled cells and empty cells were kept the same by controlling the size of the cells. By
analyzing the six sides of each hexagon, we could obtain and compare the statistical topological properties of
high-density and low-density regions of the universe in order to have a better understanding of the evolution of
the universe. We applied this hexagonal method to Two Micron All Sky Survey data and discovered significant
topological differences between the high-density and low-density regions. Both regions had significant ($>5\sigma$)
topological shifts from both the binomial distribution and the random distribution.

\end{abstract}


\section{Introduction}

Neyman et al.(1953), using statistical techniques based on counts in cells, showed that the galaxy distribution is statistically clustered rather than statistically uniform. Abell (1958) identified 2712 operationally well-defined rich clusters of galaxies from the Palomar Sky Survey plates. To better describe the clustering property, the two-point correlation function (Peebles 1980) and continual high-order correlation functions (Croton et al. 2004) are widely used. High-density regions such as filaments (Wu et al. 2012) and superclusters (Connolly et al. 1996) and low-density regions such as voids (Pan et al. 2012) have been discovered and well discussed. At the same time, people began to have an interest in the comparison of high-density and low-density regions. If the early universe grew out of quantum fluctuations, then positive and negative fluctuations will have the same probability just as a sponge-like structure (high-density regions with the same topology as low-density regions) is doomed (Gott et al. 1986).With evolution dominated by gravitational instability, the sponge-like structure is most likely the remnants of whatever growth that has either stayed in the linear regime or has damped (Bond \& Szalay 1983).

Different cells have been applied to analyze the galaxy distribution topology such as truncated octahedrons (Coxeter 1937; Gott 1967; Gott et al. 1986), Voronoi foams (Icke \& Van de Weygaert 1987), and dodecahedron cells (Kiang et al.2004). The two-dimensional (2D) topological structure of the observed universe has been investigated by a variety of methods such as the genus method (Davis \& Coles 1993; Colley 1997; Hoyle et al. 2002a, 2002b) and the percolation and filamentarity methods (Pandey \& Bharadwaj 2005). Colley (1997) found that the 2D genus of the Las Campanas Redshift Survey is very close to a Gaussian random field. Hoyle claimed the genus of the 2dF
Galaxy Redshift Survey (2DFGRS) and the Sloan Digital Sky Survey (SDSS) revealed a smaller number of isolated voids than clusters (Hoyle et al. 2002a, 2002b).

Gott et al. (1986) found that both high-density and low-density regions have similar properties in their measured density contours such as a sponge-like shape and connectedness. However, Gott et al. (1986) mentioned that a test, more direct than just measuring density contours, is needed to compare the high-density and low-density regions. Here we used the hexagonal-cell method to investigate the equivalence between high-density and low-density regions. Our hexagonal-cell method is inherited
from the dodecahedron-cell method (Kiang 2003; Kiang et al. 2004), which uses rhombic dodecahedrons as grid cells to 3D distribution of galaxies; our hexagonal method is actually a 2D application of dodecahedron cells. The use of a polyhedron/polygon here has three merits. First, it is similar to the counts-in-cells method (Efstathiou 1995), which contains statistical information about high- and low-density regions and describes the large-scale cosmic structure with a simple yet powerful technique. Second, it fully utilizes the multiple faces/sides of the polyhedron/polygon and could provide extra information about its geometry and topology. Last, but most important, for discrete-point distributions, most methods！such as the two-point correlation function, Voronoi vertices (Van de Weygaert \& Icke 1989), etc.！were not able to investigate both high-density and low-density regions equally.

The motivation of this paper is to investigate the topological relationship between low-density regions (such as voids) and high-density regions (such as clusters) in order to get a better understanding of the early universe. Our article is organized as follows. In Section 2, we present the hexagon-cells method. In Section 3, we study data from the Two Micron All Sky Survey (2MASS). Then we summarize our results in Section 4 and present our conclusions in Section 5.

\section{Hexagon cells}
If we partition the 2D space into cells with the shape of a hexagon, we obtain a hexagonal lattice, which is one of the five 2D lattice types. Furthermore, it is the only lattice that can fill the 2D space inseparably from the edges as compared to a square lattice. In the natural world, a honeycomb is an interesting example of a hexagonal lattice. For bees, a hexagonal lattice is the most efficient structure both in which to live and for storing food. Hexagonal structures also provide maximum strength; hence, they are used in designing airplane wings and satellite walls.

The reason for using hexagonal cells in this paper is that we want to have as many direction detections as possible for our unit cells. Hexagonal cells can detect their nearest neighbors in six directions and provide rich topological information. Additionally, they fill in the spaces without overlapping or missing any regions.

\subsection{Filled and empty cells}
For an application of the hexagon method on galaxy distribution, we follow a prescription analog approach taken by Kiang(2003) for rhombic cells. If a hexagonal cell contains no galaxies, then we call it an empty cell; otherwise, it is a filled cell. Empty cells therefore represent low-density regions and filled cells represent high-density regions. By these definitions, we can determine if the filled and empty regions have the same morphology in relation to each other and, in addition, provide a quantitative description of their topology.

\subsection{The number of like neighbors $n_l$}
We know that hexagons have six sides and six edges shared with neighbors. If two neighboring cells are either both filled or both empty, then we define their common edge as an inner edge; otherwise, we call it an outer edge. We use $n_l$ to define the number of inner edges. Obviously, $n_l$ is from 0 to 6, as a hexagon has six edges. The physical application of $n_l$ is to find the same kind of neighbors for each hexagon and investigate the aggregation pattern for filled and empty cells (see Figure 1 for examples of $n_l$). In this way, $n_l$ is an approximation of the two-point correlation function on a specific small scale. If filled cells are not distributed in space randomly, we can expect the distribution of $n_l$ to depart from the well-known binomial distribution; the same will be true for empty cells by symmetry. This, in turn, will tell us the morphological similarities and differences between high-density and low-density regions. Mathematically, for a given cell structure, the different $n_l$ values can be regarded as a set of base functions as they are complete and independent of each other:

\begin{equation}\label{formula1}
N_{n_l}=\sum_{i=0}^6 c_i n_i
\end{equation}

where $N_{n_l}$ is the distribution of all $n_l$ values and $c_i$ is the number of hexagons which have value ni (from $n_0$ to $n_6$ corresponding to 0 and 6 inner walls, respectively). See Figure 1 for the examples of the $n_l$ distribution ($N_{n_l}$).

\subsection{The Topological Type $\tau(m_1,m_2)$}
Based on the definition of the inner and outer edges in Section 2.2, we could divide the six edges into a group of inner edges by which we consider the connectedness of the inner or outer edges. We use $m_1$ to represent the number of connected inner edge groups and $m_2$ to represent the number of connected outer edge group; we denote the topological type by $\tau(m_1,m_2)$. For example (see Figure 2), $\tau(0,1)$ represents a single cell surrounded by six unlike neighbors (same as $n_1=0$); $\tau(1,0)$ represents a single cell surrounded by six like neighbors (same as $n_1=6$); $\tau(1,1)$ likely means that the cell is on the boundary of the clusters or voids, so its two kinds of neighbors are connected separately; $\tau(2,2)$ represents a string-like structure in which the string penetrates the cell and results in two groups for each type of neighbor; $\tau(3,3)$ is complicated and its special shape possibly implies a mesh structure.


\begin{figure}
\plotone{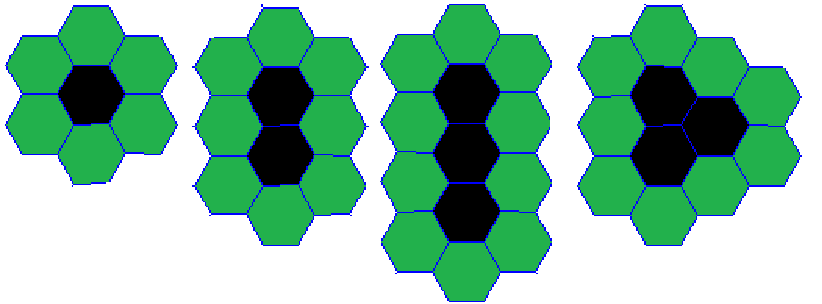}
\caption{ Filled cells are black and empty cells are green. For the filled cells:
(a) $N_{n_l}=1*n_{0}$ because the filled cell has no filled neighbors so the number of
inner walls is 0, so there is one hexagon having $n_0$ component; (b) $N_{n_l}= 2*n_1$;(c) $N_{n_l}= 2*n_1 + 1*n_2$; (d) $N_{n_l}= 3*n_2$. For the empty cells: (a) $N_{n_l}= 6*n_1$;(b) $N_{n_l}= 8*n_1$; (c) $N_{n_l}= 10*n_1$; (d) $N_{n_l}= 9*n_1$.(A color version of this figure is available in the online journal.)}
\end{figure}
%

\begin{figure}
\plotone{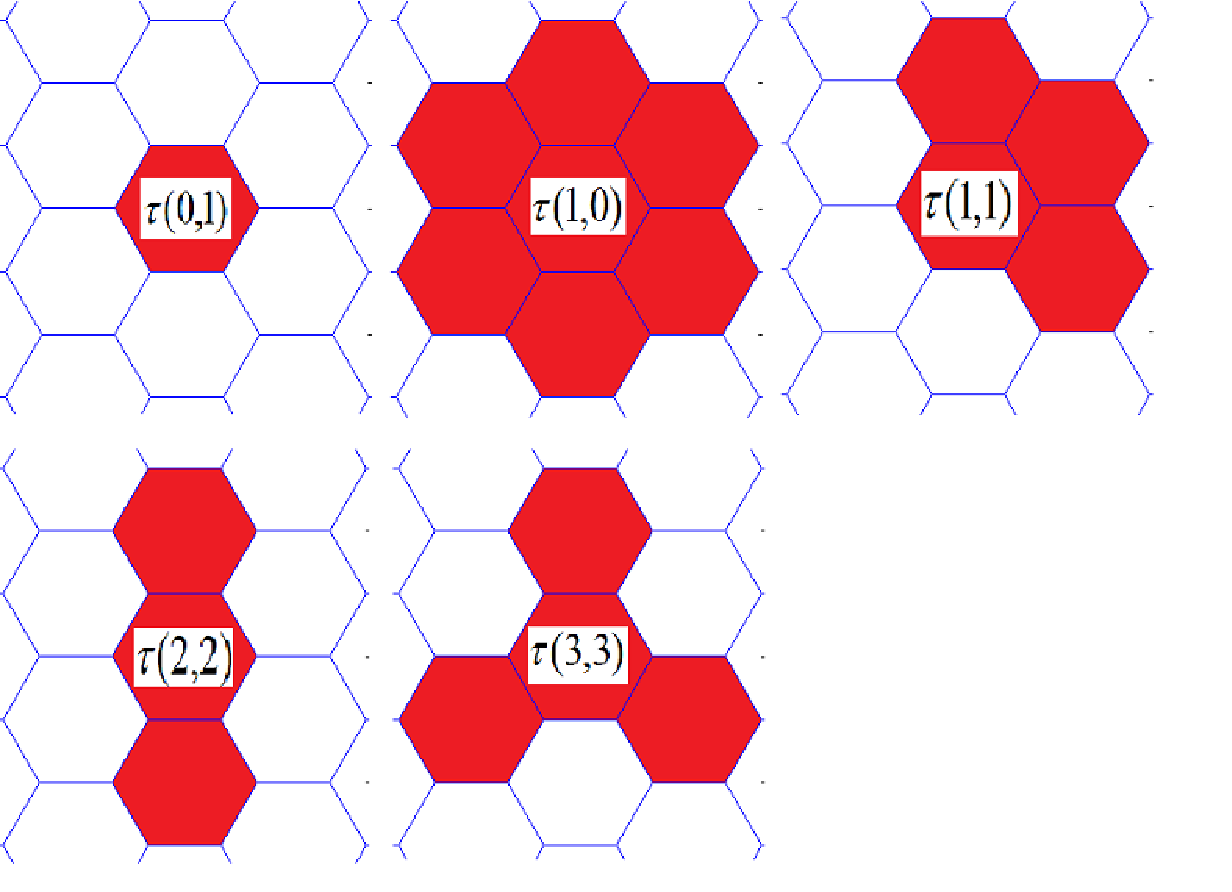}
\caption{$\tau(m_1,m_2)$ examples. Red hexagons have the same property (empty or filled) as the center-selected hexagons, whereas white hexagons have the
opposite property. For example, $\tau(3,3)$ means red hexagons are separated into three groups and white hexagons are also separated into three groups. Here we
only show one particular configuration of many possible configurations of c and $\tau(2,2)$.
(A color version of this figure is available in the online journal.)}
\end{figure}

Similarly we can also define

$$N_{\tau}=\sum_{i=0}^4 c_i \tau_i $$

Where $N_\tau$ is the distribution of $\tau(m_1,m_2)$ and $\tau_i$  represents different $\tau(m_1,m_2)$ .

Figure 2 tells us the deduction of $\tau(m,n)$. Interestingly, if we know $\tau(m,n)$, we can statistically get some hints about the structure of the map apart from just the aggregation property,e.g., whether small clumps ($\tau(0,1)$) or large clumps($\tau(1,0)$) are a majority.

\subsection{The  $\chi$ - type Indices }
To better understand the different distributions of $n_1$ and $\tau(m_1,m_2)$ between the observed and random samples, we use a $\chi$-statistic similar to that used by Kiang (2003):

\begin{equation}\label{formula2}
\chi = (O - R) / \sqrt{R}
\end{equation}

Where $O$ is for the observed sample, $R$ is for the random sample, and $\chi$  measures the degree by which the observed frequency of cells exceeds its random expectation for each different structure defined by $n_1$  and $\tau(m_1,m_2)$.

\subsection{ Zero offsets of the grid of cells}
To investigate the uncertainties of the results, we use the same method that Kiang et al. (2004) used in Rhombic Cell Analysis. Theoretically, hexagon-cell analysis is highly sensitive to the exact placing of the grid of hexagonal cells. If the lattice is displaced relative to the galaxies, then the galaxies will lie in different cells, which we call the zero offset. We start our calculations by placing the center of our zeroth cell (0, 0, 0) at Galactic coordinates (0, 0, 0). What we found was that if we displace our entire grid of cells by an amount up to and including one unit of $a_0$, in any combination of the three directions, then the resulting $n_1$ and $c$ will generally be different, but just slightly. We define the length of each cell as $1$ and simply consider four shifts: (0, 0.5), (-0.5, 0), (0.5, 0), (0.5, 0.5). The four shifts plus the original one will generate five lattices; they will be combined so we can calculate average $n_1$ and $\tau(m_1, m_2)$ values with standard deviations. Generally, the shift will result in an unequal number of filled and empty cells, but the discrepancy is very tiny (around 0.001). Moreover, it makes more sense for us to keep the length of the cell fixed rather than adjusting it to get equal numbers of filled and empty cells, so we fix the lengths of the cells when we shift them.

\subsection{Validity analysis}
The validity analysis focuses on testing the hexagonal method with the binomial distribution and Gaussian distribution. If the hexagonal method has enough resolution for recognizing the binomial and Gaussian distribution, then we expect that it is suitable as a tool for recognizing the observed galaxy distribution.

\subsubsection{ Binomial distribution theoretical analysis and simulations}
The binomial distribution gives the discrete probability distribution, $P_p (n|N)$, of obtaining exactly n successes out of $N$ Bernoulli trials！where the result of each Bernoulli trial is true with probability $p$ and false with probability $q=1-p$, as given by

\begin{equation}\label{formula3}
P_p(n|N)=\frac{N!}{n!(N-n)!} p^n q^{N-n}
\end{equation}

Since, for any neighbors of a cell, the probability of a filled or empty cell is the same (we have the same number of filled and empty cells), $p=0.5$ and $q=0.5$. For an $n_l$ (Section 2.2) analysis, we have $N=6$ (six walls for a hexagon) and n = (0, 1, 2, 4, 5, 6). Therefore according to Equation (\ref{formula3}), the probability
function $P_p(n|N)$ = (1/64, 6/64, 15/64, 20/64, 15/64, 6/64, 1/64) for n = (0, 1, 2, 3, 4, 5, 6) and $P_p(n|N)$ = (1/64, 1/64, 30/64, 30/64, 2/64) for $\tau(m_1,m_2)$ = ($\tau(0,1)$ ,$\tau(0,1)$ ,$\tau(1,0)$ ,$\tau(1,1)$ ,$\tau(2,2)$ ,$\tau(3,3)$ ); see the Appendix for this calculation.

We use a $128\times128$ random sample (named "random") to test the above theoretical results. Moreover, we also use Monte Carlo simulations to test our method. Two $128\times128$ Gaussian random samples with different power spectrums are made (one is named "G-16", presenting structures on a large scale of $2\pi L /16$ ($L$ is the size of the simulation box); another is called "G-128", corresponding to structures on a small scale of $2\pi L /128$). The motivation for using a simulation test is to check whether the hexagonal method can precisely describe the given point distribution. The samples are shown in Figure 3 and the results are shown in Figure 4. A two-point correlation function analysis is also presented！see Figure 5 for a sketch.

\begin{figure}
\plotone{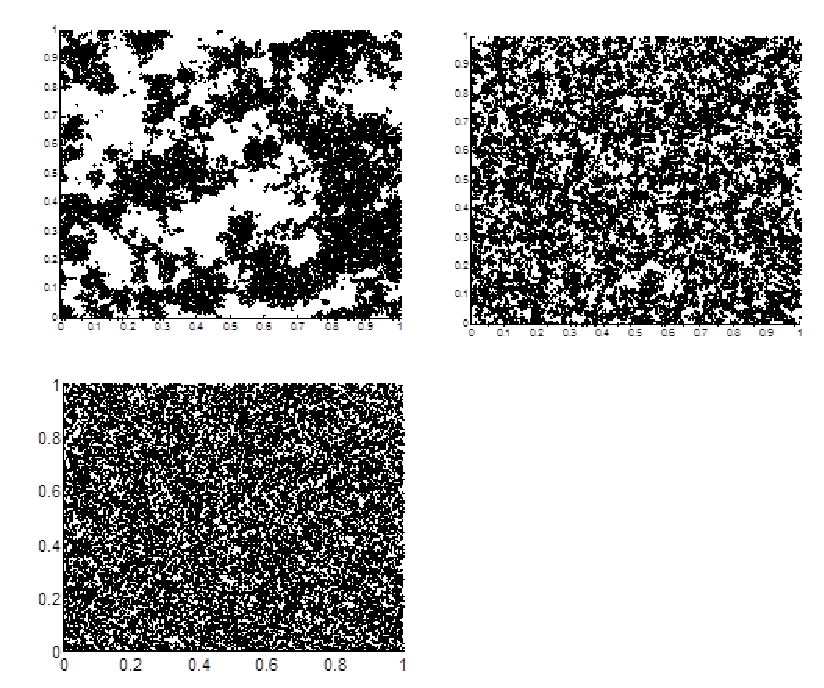}
\caption{ The sketch of three test samples.}
\end{figure}

\begin{figure}[htbp]
\centering
\mbox{
\subfigure[]{\includegraphics[scale=0.3]{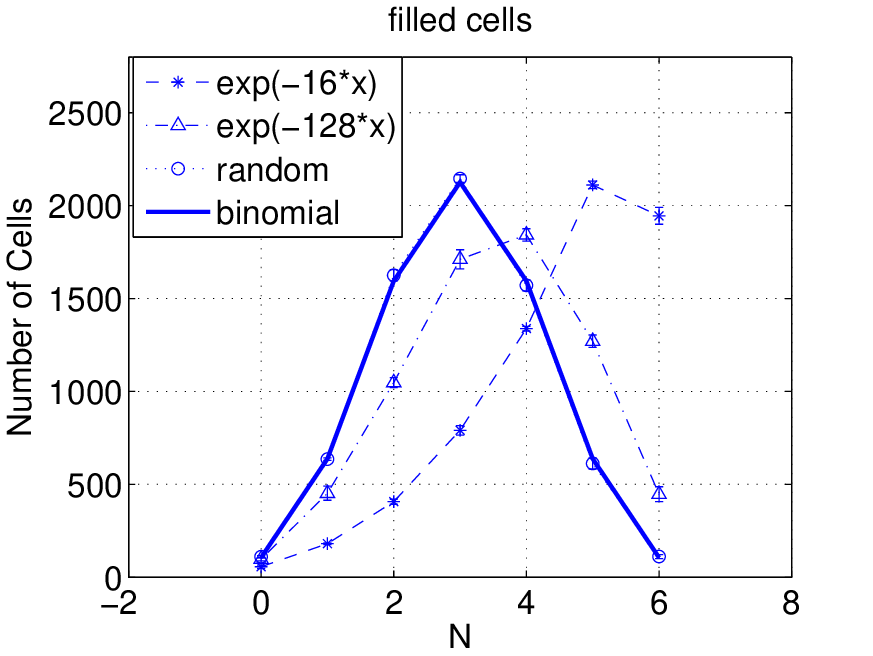}}\quad
\subfigure[]{\includegraphics[scale=0.3]{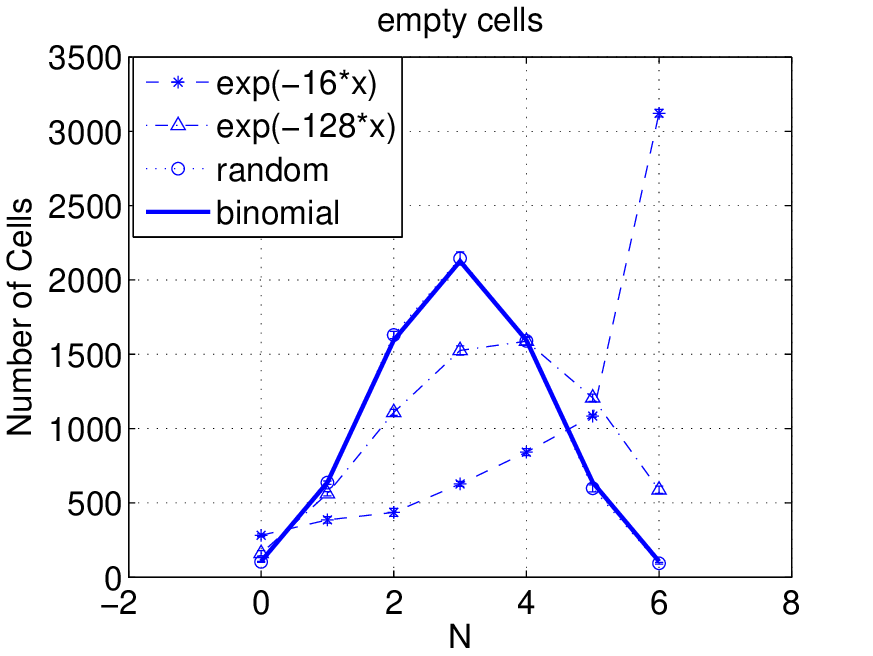}}
}
\mbox{
\subfigure[]{\includegraphics[scale=0.3]{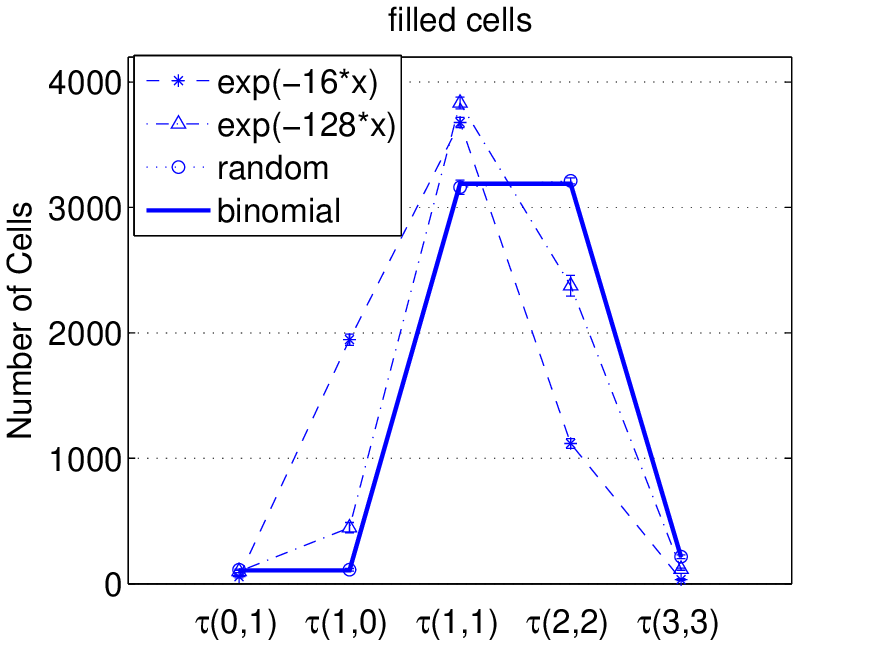}} \quad
\subfigure[]{\includegraphics[scale=0.3]{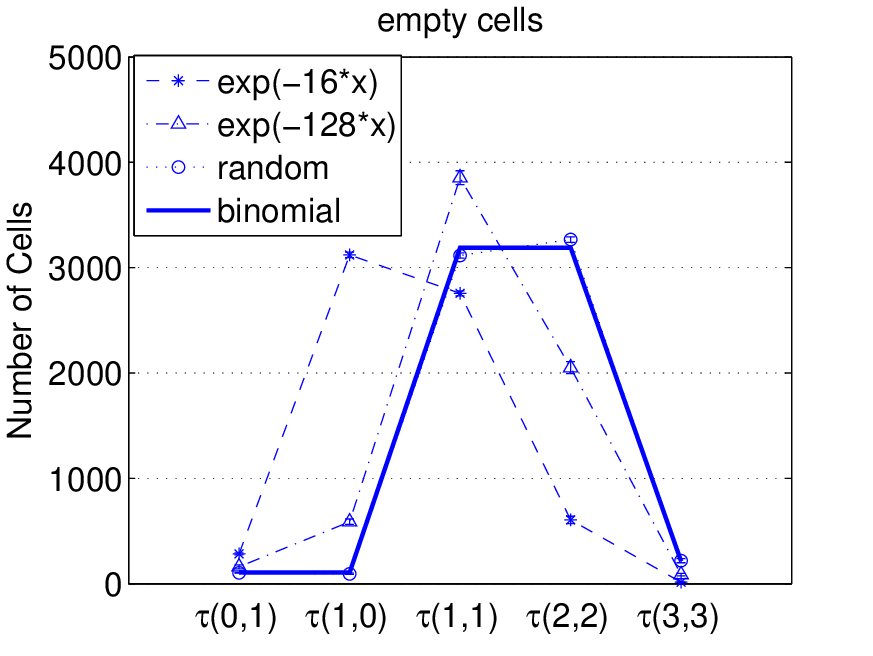}}
}
\caption{ $n_1$ and $\tau(m_1,m_2)$ results of three test samples and theoretical binomial results. The error bars are from Zero Offset and calculated from five shift samples, even though they are very small and almost undistinguishable from the figures. The random results almost overlap with the theoretical binomial results as we expected. (A color version of this figure is available in the online journal.)}
\label{fig:Fig4}
\end{figure}

From Figure 4, we find that the random sample results match the theoretical binomial results very well. However, two Gaussian samples are very different from the binomial results and also different from each other. This implies that the hexagonal method can distinguish between the different samples very well.

\subsubsection{Voronoi tessellation test}
It is important to compare the hexagonal method with other topological-analysis tools. Even though we could not find similar methods to investigate high-density and low-density regions equivalently, we still chose the Voronoi tessellation, as it is popular for statistical research in point distribution. Other methods, such as Genus, though also popular in the topological-analysis field, have to be done on a continual density field (our hexagonal method is based on a discrete density field). For
two Gaussian simulations listed in Section 2.6.1, the Voronoi tessellation can also distinguish between the clustering property of simulations very well on a large scale (see Figures 5 and 6 and Table1 for the sketches and statistical results).

From the above results, we can clearly see that the two Gaussian random samples are very different on a small scale. This coincides with the $n_1$ results of filled cells in Figure 4, as both results indicate the aggregation properties of galaxies. Moreover, the hexagonal method can give us detailed topological information about aggregation, without the smoothed density field, as shown in Figure 7.

\begin{table}[h]
\begin{center}
\caption{Results of three samples in Figure 6}\label{tab:a}
\begin{tabular}{|c|c|c|c|}
\hline
Per polygon	 & G-16	    & G-128     &	random     \\
\hline
vertices	 & 5.76	    & 5.81	    & 5.99         \\
\hline
area	     & 5.89e-5	& 6.07e-5	& 6.08e-5      \\
\hline
perimeter	 & 0.0279	& 0.0305	& 0.0312       \\
\hline
side length	 & 0.0048	& 0.0052	& 0.0052       \\
\hline
\end{tabular}
\end{center}
\end{table}

\begin{figure}[htbp]
\epsscale{1.0}
\plotone{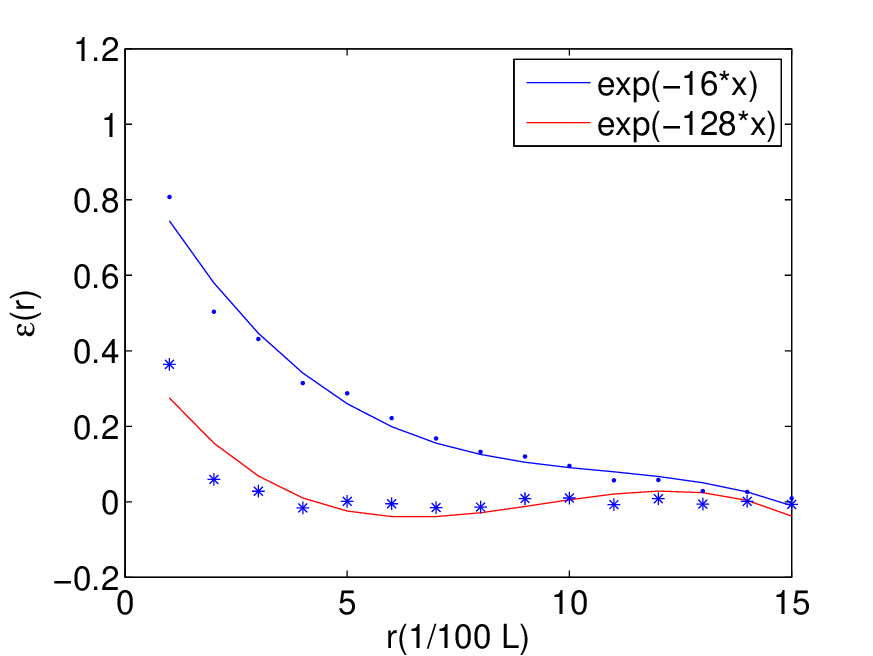}
\caption{Two-points correlation function resulting of two Gaussian-random simulation function results}
\end{figure}

\begin{figure}[htbp]
\epsscale{1.0}
\plotone{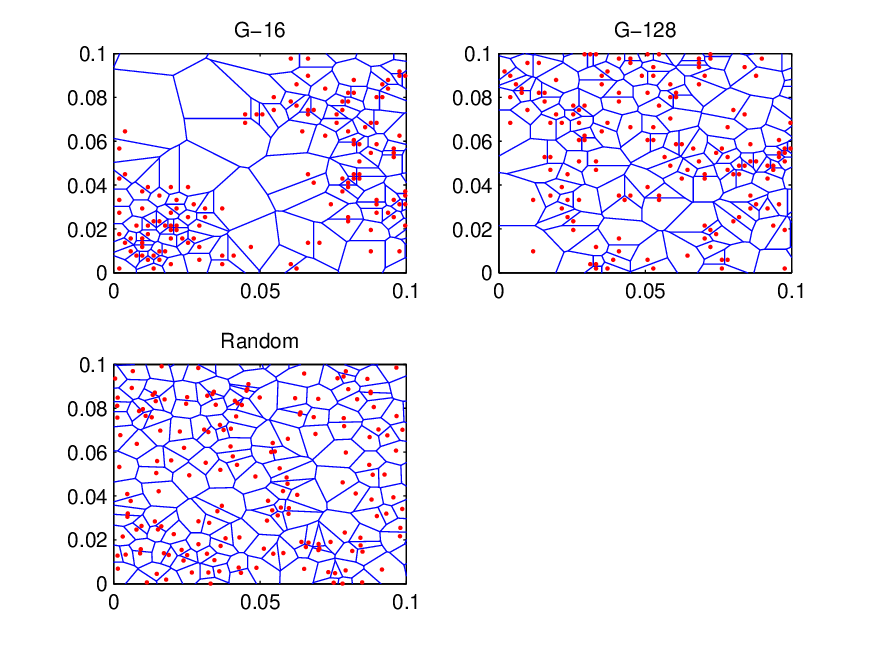}
\caption{An example of Voronoi tessellation result of three test samples (only show a small area for the entire sample).}
\end{figure}

\section{2MASS Data}
The 2MASS Extended Source Catalog (XSC; Jarrett et al.2000) data we used were provided by Cao et al. (2006); detailed information can be found in their paper. Simply speaking, they selected galaxies inside an elliptical isophote with a surface brightness of 20 mag $arcsec^2$ in the Ks band. To have 2D plane data, they transformed the 2MASS data to an equal-area projection via the Lambert azimuthal algorithm (see Figure 8):

\begin{equation}\label{formula4}
\begin{array}{l}
x_1 = R \sqrt{2-2| sin(b)|} cos (l)\\
x_2 = R \sqrt{2-2| sin(b)|} sin (l)
\end{array}
\end{equation}

\begin{figure}[htbp]
\centering
\mbox{
\subfigure[]{\includegraphics[scale=0.3]{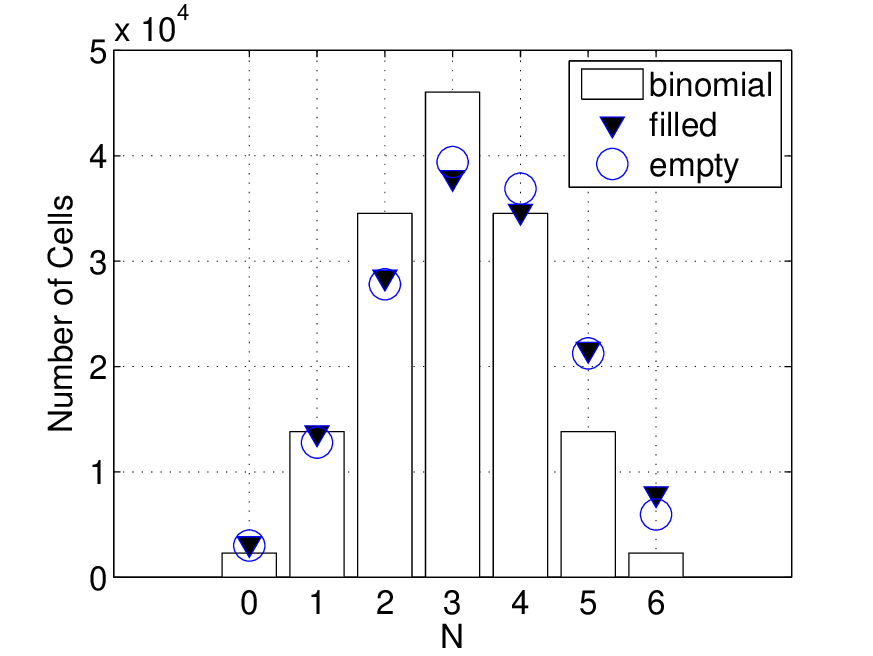}}\quad
\subfigure[]{\includegraphics[scale=0.3]{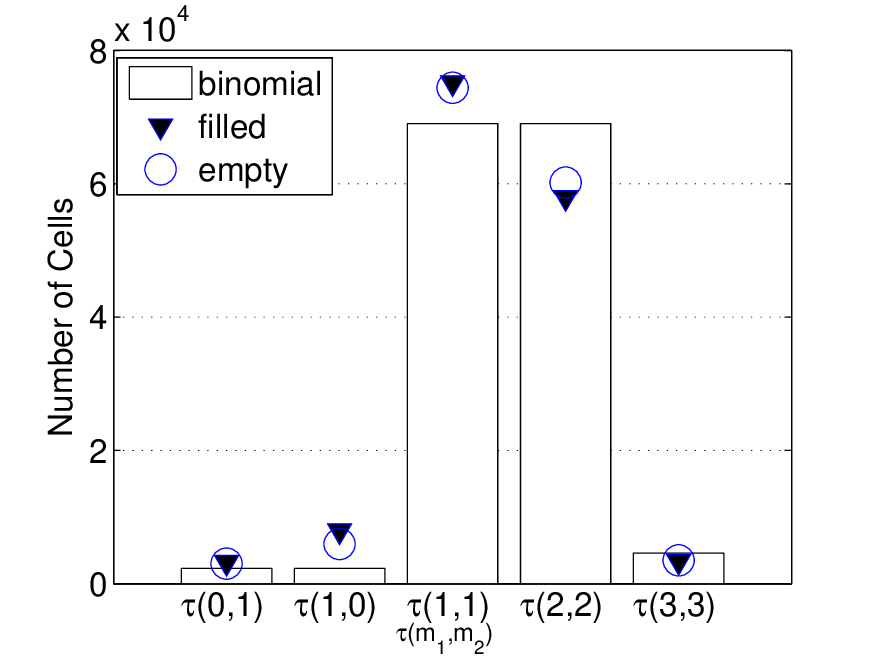}}
}
\caption{$n_1$ and $\tau(m_1,m_2)$ results of the observed sample. Triangles represent the filled cells and circles represent the empty cells. The numbers of filled and empty cells are kept the same by adjusting the sizes of the cells. We used 354,560 hexagons. This is the average value of 10 observed results from the zero offsets of grid cells mentioned in Section 2.5. The $\sigma$'s are extremely small！because we used a huge number of hexagons and galaxies, the standard deviation is around 50 to 300 for all $n_1$  and $\tau(m_1,m_2)$ ！and cannot be displayed in the figure. }
\label{fig:Fig7}
\end{figure}

where $R$ is a relative scale factor, $b$ is the Galactic latitude, and $l$ is the Galactic longitude. This hemisphere scheme projects the whole sky into two circular planes, a northern and southern sky. We select a square with $123^\circ.88 \times 123^\circ.88$ in the central part of each circular plane.We have two fields of $123^\circ.88 \times 123^\circ.88$ in the northern and southern skies. The northern square has 323,000 galaxies and the southern includes 342,000 galaxies.

\section{Results}
After adjusting the length of the cells and excluding the outermost cells (to avoid boundary distortion), we use 172,320 filled and 172,320 empty hexagonal cells to fill up the 2D northern sky of 2MASS, which has roughly 290,000 galaxies inside used hexagons. 

The observed sample results of $n_1$  and $\tau(m_1,m_2)$ are shown in Figure 7.

\begin{figure}[htbp]
\epsscale{1.0}
\plotone{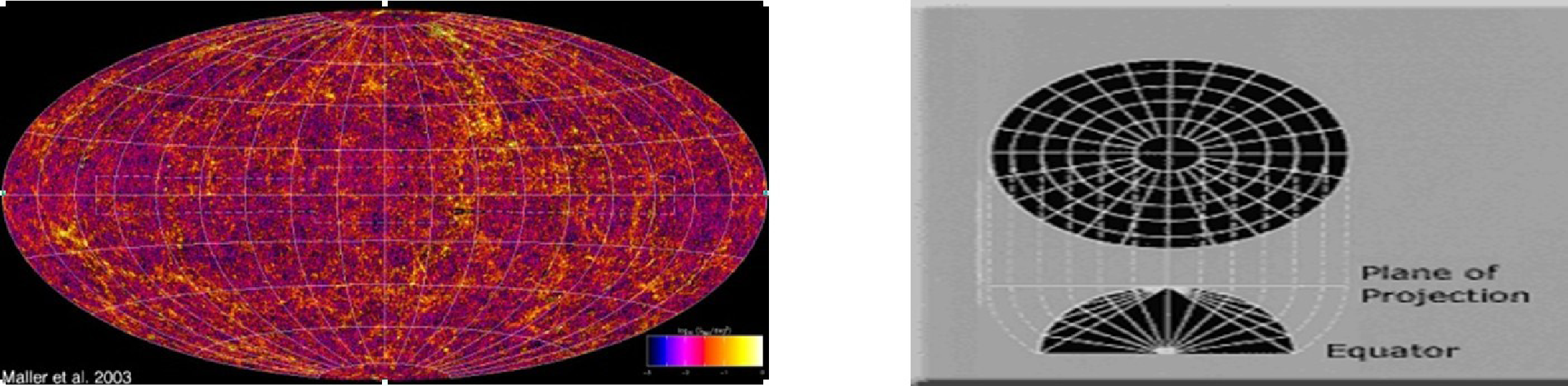}
\caption{2MASS data and projection scheme.\protect\footnotemark[1] }
\end{figure}

\footnotetext{http://egsc.usgs.gov/isb/pubs/MapProjections/projections.html}
From Figure 7, the filled and empty hexagon cells appear to have very close $n_1$  and $\tau(m_1,m_2)$ distributions, but considering the extremely small standard deviations, they are actually quite different (the reader may inquire whether the Zero Offset technique could be used to get an appropriate standard deviation, but
using a set of random samples with different seeds would give an even smaller standard deviation for our $n_1$  and $\tau(m_1,m_2)$ distribution). Table \ref{tab:a} shows the $\chi$ results mentioned in Equation (\ref{formula2}) with corresponding standard deviations from 10 Zero Offset grids. Note that here we use a binomial distribution as a reference instead of comparing filled and empty cells directly; this is because we try to use binomial distributions as a "ruler" to give the difference between filled and empty cells a more physical meaning.


\begin{table}[h]
\begin{center}
\caption{$\chi$ distance to binomial distribution (northern sky)}\label{tab:a}
\begin{tabular}{|c|c|c|c|c|}
\hline
         & Filled           & Standard         & Empty           & Standard  \\
         & cells            & deviation        & cell            & deviation \\
         & $\chi_{filled} $ & $\chi_{filled} $ & $\chi_{empty} $ & $\chi_{empty} $ \\
\hline
$n_1=0$  & 18.6             & 0.97             & 14.7            & 1.22       \\
\hline
$n_1=1$  & -1.0             & 0.66             & -8.8            & 0.34       \\
\hline
$n_1=2$  &-32.6             & 0.68             &-36.2            & 0.86       \\
\hline
$n_1=3$  &-37.7             & 0.87             &-30.9            & 0.27       \\
\hline
$n_1=4$  & 1.1              & 1.13             & 12.7            & 0.72       \\
\hline
$n_1=5$  & 66.4             & 1.44             & 63.3            & 1.11       \\
\hline
$n_1=6$  &117.1             & 1.97             & 76.1            & 0.85       \\
\hline

$\tau(0,1)$ & 18.6          & 0.97             & 14.7            & 1.22       \\
\hline
$\tau(1,0)$ &117.1          & 1.97             & 76.1            & 0.85       \\
\hline
$\tau(1,1)$  & 23.2         & 0.99             & 20.4            & 0.63       \\
\hline
$\tau(2,2)$  & -42.3        & 0.50             & -33.7           & 0.74       \\
\hline
$\tau(3,3)$ &-18.7          & 1.00             & -16.3           & 1.17       \\
\hline
\end{tabular}
\end{center}
\end{table}

The same method is applied on the 2MASS southern sky map and gives similar results, shown in Table \ref{tab:b}. Note that here we cannot combine 2MASS northern and southern results together because there is a slight structure difference between the two sky maps. For example, $\tau(0,1)$ is 105.8 in the southern sky map but 117.1 in the northern sky map; this is a $\sim5\sigma$ difference and cannot be neglected.


\begin{table}[h]
\begin{center}
\caption{$\chi$ distance to binomial distribution (southern sky)}\label{tab:b}
\begin{tabular}{|c|c|c|c|c|}
\hline
         & Filled           & Standard         & Empty           & Standard  \\
         & cells            & deviation        & cell            & deviation \\
         & $\chi_{filled} $ & $\chi_{filled} $ & $\chi_{empty} $ & $\chi_{empty} $ \\
\hline
$n_1=0$  & 18.5             & 1.27             & 12.2            & 0.80       \\
\hline
$n_1=1$  & -1.9             & 1.11             & -8.7            & 1.18       \\
\hline
$n_1=2$  &-31.4             & 0.96             &-34.0            & 1.808      \\
\hline
$n_1=3$  &-35.5             & 0.57             &-29.5            & 0.55       \\
\hline
$n_1=4$  & 1.7              & 1.13             & 12.1            & 1.22       \\
\hline
$n_1=5$  & 62.3             & 1.42             & 62.5            & 1.62       \\
\hline
$n_1=6$  &105.8             & 1.93             & 74.0            & 1.40       \\
\hline

$\tau(0,1)$ & 18.5          & 1.27             & 12.2            & 0.80       \\
\hline
$\tau(1,0)$ &105.8          & 1.93             & 74.0            & 1.40       \\
\hline
$\tau(1,1)$  & 22.7         & 1.05             & 21.3            & 0.92       \\
\hline
$\tau(2,2)$  & -41.0        & 1.03             & -32.8           & 0.79       \\
\hline
$\tau(3,3)$ &-18.2          & 0.68             & -15.1           & 0.5       \\
\hline
\end{tabular}
\end{center}
\end{table}

Note in Table \ref{tab:a} we see that $n_1$=0 and $\tau(0,1)$ are identical as are  $n_1$=6 and $\tau(1,0)$; this is consistent with what we describe in Section 2.3.

We see that the results of the $n_1$=1, 3, 4, 6 and $\tau(0,1)$, $\tau(2,2)$ distributions have significant differences between the filled and empty cells. We also note the following results from Table \ref{tab:a}.

$\bullet $	From $n_1$=6 (or $\tau(1,0)$ ), filled cells have a much stronger (21$\sigma$ for the northern sky and 15$\sigma$ for the southern sky) aggregation property than empty cells. This is consistent with Hoyle et al. (2002a, 2002b) who found that the number of isolated voids is smaller than the number of clusters in 2DFGRS.

$\bullet$ 	The most general case is n = 3 (20/64 probability for a standard distribution); however, we can see the $n_1$ value at n = 3 is smaller than the standard distribution. This can be explained as aggregation causing the filled and empty cells to have a greater probability in $n_1>3$ than the standard distribution, hence reducing the probability at n = 3.

$\bullet$   For $n_1$=1, filled cells are closer (12$\sigma$ for the northern sky and 6$\sigma$ for the southern sky) to the standard distribution than empty cells, which implies that empty cells have either fewer or longer fiber structures than filled cells, as the end of a fiber always has $n_1$=1. Since $n_1$=1 must result in a $\tau(1, 1)$ structure, we do not see much difference between the filled and empty cells for $\tau(1, 1)$ in Table \ref{tab:c}. This suggests that the $n_1$=1 structure is a minority in most $\tau(1, 1)$ structures. In other words, $\tau(1, 1)$ includes many more cells on the boundary of filled and empty cells than those on the end.

$\bullet$  	$\tau(2, 2)$ also has a significant difference (12$\sigma$ for the northern sky and 8$\sigma$ for the southern sky) between filled and empty cells. Empty cells are much more frequently penetrated by filled cells than filled cells are penetrated by empty cells, which implies that empty cells have less fiber structure than filled cells (consistent with the $n_1$=1 results).

\section{Conclusions}
With a direct measurement by the hexagonal-cell method, we found there are significant differences in topology between high density regions and low-density regions in 2MASS data. The accuracy is assured by the Zero Offset technique.We found that low-density regions have significantly less fiber structure and that they are similarly less clustered than high-density regions. These differences could be caused by evolution, which suggests that evolution might include more things than gravitational
instability (Bond \& Szalay 1983) or be caused by the original asymmetry of quantum fluctuation in the early universe (Turner 1999). Overall, low-density regions have a slightly closer topology, in relation to a random distribution, than high-density regions.We also found that the topology of the 2MASS southern sky map is slightly closer to a random distribution than the northern sky.Ourwork provides a supplement to genus analysis, which shows that high-density and low-density regions have almost identical properties (Park et al. 2001).

For a given binomial distribution, Section 2.6.1 describes a precise analytical approach. For other given distributions, as long as we can analytically define the distribution, we should be able to find an analytical approach. Further research is needed, which we will discuss in future papers.

~\\

\appendix 

Here we present the $\tau(m_1,m_2)$ calculation for a binomial distribution. Since there are six walls, and each wall could be either an inner or outer wall, the total number of combinations gives N=$2^6$=64 arrangements. If we use "1" for an inner wall and "0" for an outer wall, we can write the numbers 0 to 63 in a
6-digit binary format, such as 000000, 000001, etc. Then the 64 binaries represent all possible arrangements of $\tau(m_1,m_2)$. We list the results in Table 4.


\begin{table}[h]
\begin{center}
\caption{the binomial probability function for $\tau(m_1,m_2)$ }\label{tab:c}
\begin{tabular}{|c|c|c|c|c|c|c|c|}
\hline
            & count  & \multicolumn{5}{|c|}{ arrangements } \\
\hline
$\tau(0,1)$ & 1      & \multicolumn{5}{|l|}{ 000000 } \\
\hline
$\tau(1,0)$ & 1      & \multicolumn{5}{|l|}{ 111111 } \\
\hline
$\tau(1,1)$ & 30     & \multicolumn{5}{|c|}{000001 000010 000011 000100 000110 000111 } \\
            &        & \multicolumn{5}{|c|}{001110 001111 010000 011000 011100 011110 } \\
            &        & \multicolumn{5}{|c|}{100001 100011 100111 101111 110000 110001 } \\
            &        & \multicolumn{5}{|c|}{001000 001100 011111 100000 110011 110111 } \\
            &        & \multicolumn{5}{|l|}{111000 111001 111011 111100 111101 111110 } \\
\hline
$\tau(2,2)$ & 30     & \multicolumn{5}{|c|}{000101 001001 010100 001011 001101 010001 } \\
            &        & \multicolumn{5}{|c|}{010100 010110 010111 011001 011010 011011 } \\
            &        & \multicolumn{5}{|c|}{100100 100101 100110 101000 101001 101011 } \\
            &        & \multicolumn{5}{|c|}{010010 010011 011101 100010 101100 101101 } \\
            &        & \multicolumn{5}{|l|}{101110 110010 110100 110101 110110 111010 } \\
\hline
$\tau(3,3)$ & 2      & \multicolumn{5}{|l|}{010101 101010                             } \\
\hline
\end{tabular}
\end{center}
\end{table}

In conclusion, the binomial probability function for $\tau(m_1,m_2)$ is 1/64, 1/64, 30/64, 30/64, 2/64 for $\tau(0,1)$, $\tau(1,0)$,$\tau(1,1)$,$\tau(2,2)$, $\tau(3,3)$.



\clearpage


\end{document}